\documentclass[12pt]{iopart}
\usepackage{graphicx}

\def\degrees{270}

\newcommand{\be}{\begin{equation}} \newcommand{\ee}{\end{equation}}
\newcommand{\gsim}{\lower.7ex\hbox{$\;\stackrel{\textstyle>}{\sim}\;$}}
\newcommand{\llsim}{\lower.7ex\hbox{$\;\stackrel{\textstyle<}{\sim}\;$}}
\newcommand{\bea}{\begin{eqnarray}} \newcommand{\eea}{\end{eqnarray}}

\begin{document}

\begin{flushright}
 IFT-UAM/CSIC-07-07
\end{flushright}

\title{Inflation in uplifted Supergravities}

\author{B.~de Carlos$^1$, J.~A.~Casas$^2$,  A.~Guarino$^2$, J.~M.~Moreno$^2$ \\ and O.~Seto$^2$}

\address{
$^1$ School of Physics and Astronomy, University of Southampton, \\ Southampton SO17 1BJ, UK \\
$^2$ Instituto de F\'{\i}sica Te\'orica, C-XVI, UAM \\
 28049, Madrid, Spain\\
}
\ead{b.de-carlos@soton.ac.uk, alberto.casas@uam.es, adolfo.guarino@uam.es, jesus.moreno@uam.es, osamu.seto@uam.es}

\begin{abstract}
We present a model of slow-roll inflation in the context of effective Supergravities arising from string theories. The uplifting of the potential (to generate dS or Minkowski vacua) is provided by the D-term associated to an anomalous U(1), in a fully consistent and gauge invariant formulation. We develop a minimal working model which incorporates eternal topological inflation and complies with observational constraints, avoiding the usual obstacles to implement successful inflation ($\eta$ problem and
initial condition problem among others).

\end{abstract}

\maketitle

\section{Introduction}

The topic of inflation coming from string models is experiencing a comeback driven by both progress in our theoretical understanding of the models and the increasing precision of observations. Whereas the former is commonly identified, in its phenomenological aspects, with the publication of the moduli stabilisation mechanism proposed by  Kachru et al. (KKLT)~\cite{Kachru:2003aw}, the recent experimental data released by the WMAP collaboration~\cite{Spergel:2006hy} give support to the inflationary picture, as well as provide us with stringent limits on some of the cosmological parameters, something that can immediately be implemented within inflationary model building.    

This revival comprises different string models where the inflaton is normally a modulus, either parametrising the distance between branes (as proposed originally by Dvali and Tye~\cite{Dvali:1998pa}), or the geometry/structure of the compactified space. This second option, also denoted as {\em modular inflation}, has been studied ever since the first models of moduli stabilisation were put forward~\cite{Binetruy:1986ss}. Inflation was not working at all because either the moduli were flat at all orders in perturbation theory or, when including non perturbative effects, their potential would be too steep to inflate~\cite{Brustein:1992nk}.  Moreover we have to add the fact that, at that stage, all models of moduli stabilisation predicted a negative vacuum energy. Despite of all these problems, a few partially successful examples were built~\cite{Copeland:1994vg,Banks:1995dp,Bailin:1998tu}.

Progress, as explained in the first paragraph, has been driven by our increasing understanding of flux compactification (for a review see \cite{Douglas:2006es}) and its crucial role in stabilising moduli. In addition, several mechanisms to solve
the long standing problem of the appearance of Anti de Sitter (AdS) vacua have been
proposed. In particular, in the context of Type IIB theory, KKLT -see also \cite{Choi:2006bh}- considered the 
presence of anti-D3 branes as the source of an additional term in the scalar potential, which breaks SUSY explicitly and may uplift the minima from AdS to dS.
In order to improve the aesthetics and the control over the effective theory, an alternative, fully supersymmetric, approach based on the presence of a non-vanishing
(Fayet-Iliopoulos) D-term as the uplifting contribution to the potential, was proposed 
in ref.~\cite{Burgess:2003ic}. In its original formulation, this mechanism had important inconsistencies, that were fixed in ref.~\cite{Achucarro:2006zf}. In that paper,
a consistent formulation of these so-called ``uplifting" D-terms in the context of $N=1$ Supergravity was established, implying the presence of chiral matter (see more details
in the next section). Explicit viable examples were given, where a $T$-modulus and 
the chiral matter get stabilised at phenomenologically relevant values and positive vacuum energy. The scenario was mainly determined by the requirements of supersymmetry and gauge  invariance and, therefore, left little room for fine-tuning the parameters entering the superpotential, which increases the predictive power. (Alternative approaches to
achieve the desired uplifting can be found in refs.~\cite{Villadoro:2005yq, Parameswaran:2006jh, Westphal:2006tn}.)

In the present article we carry on along the lines of modular inflation in this new context of uplifted Supergravities\footnote{Other, related work, in the topic of string/brane inflation is given 
here~\cite{Burgess:2001fx}--\cite{Kallosh:2007ig}.}, and propose and study a scenario based on the
setup of ref.~\cite{Achucarro:2006zf} plus an extra singlet, where suitable inflation takes place.
In order to do that we take into account all possible observational constraints which, together with the symmetries of the model, will determine the structure and size of the different couplings in the superpotential. 

The paper is organised as follows: in section~\ref{sec:rev}, we review the basics of the scenario presented in \cite{Achucarro:2006zf} as an introduction to our model of inflation. In
section~\ref{sec:etaic} we explore the chances and problems to implement inflation within this kind of framework, pointing out generic problems affecting to natural candidates to the role of inflaton. In section~\ref{sec:model} we present the model itself and show how it gives rise to successful slow roll inflation. We also discuss our main results and, in section~\ref{sec:conclu}, we conclude.

\section{Description of the scenario}
\label{sec:rev}

In this section we briefly review the scenario considered in  ref.~\cite{Achucarro:2006zf}. It describes an effective Supergravity  coming from type IIB string theory, although it can also be realised in the context of heterotic strings. Along the lines of the moduli stabilisation mechanism proposed by KKLT~\cite{Kachru:2003aw}, we assume that all moduli but one, an overall $T$-modulus, have been stabilised at a high scale due to the presence of fluxes. We also assume, as it is common in these setups, that gaugino condensation happens, with gauge group SU($N$), due to stacks of $N$ D7-branes wrapped on some 4-cycle of the Calabi-Yau space. For each SU($N$) there typically appears a U(1) factor.  Some of these U(1)s, or combinations of them, can be anomalous.  As proposed in ref.~\cite{Burgess:2003ic}, the corresponding D-terms can provide the required uplifting of the potential in order to have de Sitter vacua. This has the advantage that SUSY is not explicitly broken (as it was
by the anti D3 brane contributions originally considered in KKLT), so corrections can be more reliably computed and kept under control. However, the setup of ref.~\cite{Burgess:2003ic} had some serious
inconsistencies~\cite{Choi:2005ge, deAlwis:2005tf}, arising from the lack of gauge invariance
of the formulation. Nevertheless, as shown in  ref.~\cite{Achucarro:2006zf},
this scheme can be made gauge invariant, which imposes the presence of chiral matter transforming typically as $(N,   \bar{N})$ with abelian charges $(q, \bar{q})$, for the whole setup to be consistent (this mechanism was confirmed in explicit string constructions in ref.~\cite{Haack:2006cy}). As stressed in ref.~\cite{Achucarro:2006zf}, the constraints coming from enforcing gauge invariance are sufficiently strong to determine the form of the superpotential. Moreover, the presence of the anomalous U(1) group generates a Fayet-Iliopoulos term which enters the D part of the scalar potential and
is responsible for the uplifting.

Based on this information one can construct a simple model assuming that the symmetries in the hidden sector are dictated by a unique SU($N$) times the anomalous U(1). In its simplest possible form, the model contains
\begin{itemize}
\item A modulus, $T$.
\item Two chiral multiplets, $Q$, $\bar{Q}$, transforming as $(N,   \bar{N})$ with abelian charges $(q, \bar{q})$ (i.e. $N_f=1$ in the usual notation). 
\end{itemize}
The modulus field transforms non trivially under the U(1) group,
\begin{equation}
T \rightarrow T+ i{\delta_{GS}\over 2}\epsilon\ \ ,
\end{equation}
and the corresponding transformation of the Lagrangian compensates
the SU($N$)$^2 \times $ U$(1)_X$ and  U$(1)_X^3$ anomalies, according to the Green-Schwarz
mechanism. The resulting condition reads
\bea
\label{anom3}
\delta_{GS} =-{(q + \bar q)\over 2 \pi k_N}  = -{N  (q^3 + \bar
q^3)\over 3 \pi k_X}   \;\;,
\eea
where $k_N, k_X$ are ${\cal O}(1)$ constant factors that enter in the definition
of the corresponding gauge couplings~\cite{Ibanez22}.
It is also well known that  the theory gets strongly coupled in the infrared. As a consequence, gaugino condensation takes place at some scale, $\Lambda$, and also squark meson condensates 
\be
M^2 = 2 Q \bar Q \;\;,
\ee
are formed. A non perturbative superpotential term is generated~\cite{Taylor:1982bp,Lust:1990zi,deCarlos:1991gq},
\be
W_{\rm np} = (N-1)  \left(\frac{2\Lambda^{3N-1}}{M^2}
                             \right)^{1\over N-1} = (N-1) \left(\frac{2}{M^2}\right)^{1\over N-1}  
             {\rm e}^{-4 \pi k_N T\over N-1}
\label{Wnp}\;\;.
\ee
Note that this term is, as expected, invariant under U(1) transformations.

As usual in ${\rm N}=1$ SUGRA, the potential is the sum of an F  and a D-part,
\be
V=V_{\rm F} + V_{\rm D}\;\;.
\label{VVFVD}
\ee
Using Planck units,  $M_{\rm Planck}^{-2} = 8 \pi G_{\rm N} =1$,
 $V_{\rm F}$ is given by
\be
V_{\rm F} = {\rm e}^K \left[K_{ij}^{-1} D_i W D_j \bar{W} - 3|W|^2 \right] \;\;,
\label{pot}
\ee
where $K_{ij}^{-1}$ is the inverse K\"ahler metric, $K_{ij}= \partial^2 K/\partial \Phi_i\partial \bar \Phi_j$,
and $D_iW=K_i W + W_i$ is the K\"ahler derivative (here subindices denote derivatives with respect to all scalar fields, $\Phi_i$).  In our case, $V_{\rm F}$ can be computed using the complete superpotential, $W = W_0 + W_{\rm np} $ [where $W_0$ is the (real)
effective flux parameter] and the K\"ahler potential~\footnote{
We are assuming here a minimal K\"ahler potential for the matter fields. This is 
a simplification, but it does not affect the main results and conclusions. We will 
come back to this point in sect.~4.}
\be
\label{KIIB}
K = -3 {\rm log}(T+\bar{T})  + |Q|^2 + |\bar{Q}|^2 = -3 {\rm log}(T+\bar{T})  + |M|^2 \;.
\ee
Note that we have used $|Q|^2=|\bar{Q}|^2$, which  is dynamically dictated by the cancellation of the SU($N$) D-term (in the $N_f=1$ case). The D part of the potential associated with the U(1) is given by
\be V_{\rm D} =\frac{\pi}{2k_X (T + \bar{T})} \left((q+\bar{q}) |M|^2  -
\frac{3\delta_{GS}}{T + \bar{T} }\right)^2 \;\;,
\label{VDIIB}
\ee
which is positive definite. For reasonable values of $N, q, \bar q$ and $k_N$ it is possible to choose 
$W_0$ in such a way that {\em i)} there are minima of the F-potential with broken SUSY and negative vacuum energy; {\em ii)} $V_{\rm D}$ is non-zero and sizeable enough to uplift these vacua from AdS to dS.

As explained in ref.~\cite{Achucarro:2006zf}, the absence of physical vacua with unbroken SUSY comes from the fact that the conditions $D_T W = D_M W=0$ can only be fulfilled simultaneously in the
decompactification limit, Re($T$) $\rightarrow \infty$ . 
This can be easily checked by computing
\be
\frac{1}{K_T} D_T W-\frac{1}{K_M} D_M W = W_{\rm np} 
\left( \frac{T+\bar{T}}{3} a + \frac{b}{|M|^2} \right) \;\;,
\label{nosusy}
\ee
with $(a,b)$ defined by expressing (\ref{Wnp}) as $W_{\rm np} \propto M^{-b} {\rm e}^{-a T} $. The positiveness of the $(a,b)$ parameters imply that this equation has no solution consistent with 
$D_T W = D_M W=0$ for physical values ( $0 <$ Re$(T)< \infty $) of the modulus field.

Let us now review the main features of the vacuum structure of these models. For the remainder of this analysis it is particularly convenient  to split the complex fields $T$ and $M$ in the following way
\bea T & = & T_{\rm R}+ {\rm i} T_{\rm I} \nonumber \;, \\ 
     M & = & \rho_{\rm M} {\rm e}^{{\rm i}  \alpha_{\rm M}} \;.  \eea
As we have mentioned, there is always a SUSY conserving vacuum at $T_{\rm R} \rightarrow \infty $. 
Besides this minimum, for given values of $\rho_{\rm M}$ and $T_{\rm R}$,
the potential gets always minimal when $\alpha_{\rm M}$ and $T_{\rm I}$ 
are such that
$W_0$ and $W_{np}$ are aligned in the complex plane (for more details see the Appendix).
For real $W_0$ this translates into the condition
\be
\varphi_{\rm np} \equiv - \frac{2}{N-1} ( \alpha_{\rm M} +  2\pi k_N T_{\rm I} )  =  n\pi \;.
\label{phase}
\ee
where $\varphi_{\rm np}$ is the phase of $W_{np}$ and $n$ is even or odd depending on the 
details of the model \cite{Achucarro:2006zf}.
Actually,  $\alpha_{\rm M}$ and $T_{\rm I}$ appear
in the potential only through the $\varphi_{\rm np}$ combination, which can thus be set to its minimising value. Hence, the minimisation can be reduced to a two variable problem, namely to find the values of $(T_{\rm R}, \rho_{\rm M})$ at the minimum. 
\begin{figure}[ht]
\centering						
\includegraphics[angle=\degrees, width=12cm]{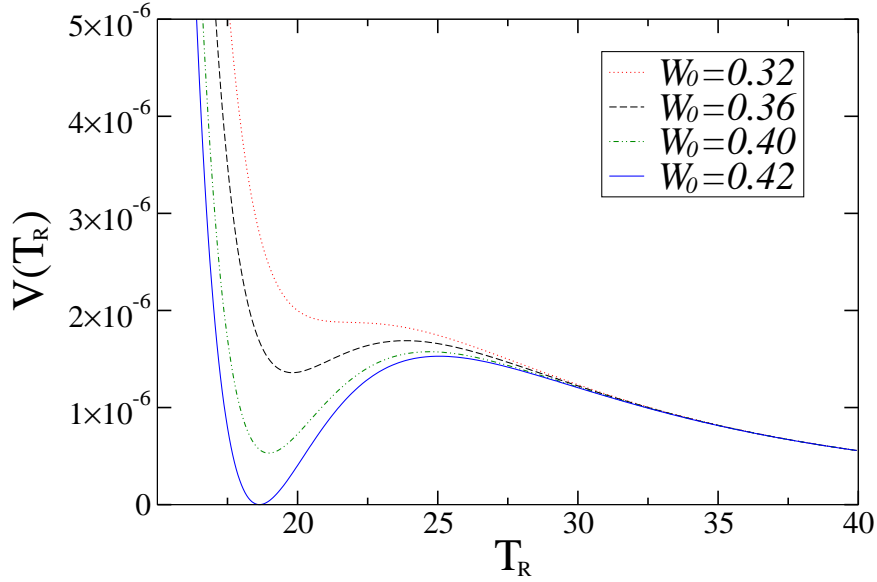}
\caption{{ Plot of the scalar potential, as a function of $T_{\rm R}$, for the example 
shown in the text and $\rho_{\rm M}$ chosen as the value minimising the potential at each
$T_{\rm R}$.}}
\label{W0}
\end{figure}

It is worth noticing that the energy of the vacuum is controlled by the size of the flux parameter $W_0$.   In fig.~\ref{W0} we plot the dependence of the scalar potential with $W_0$ as a function of $T_{\rm R}$, with the non-perturbative phase, $\varphi_{\rm np}$, fixed to its minimising value. 
The matter condensate, $\rho_{\rm M}^2$, changes along the path and is determined by the
extremal condition $\delta_{\rho_{\rm M}} V = 0 $.
The particular model is defined by $N=20$, $q=1$, $\bar{q}=1/10$, $k_N=1/2$ (the remaining parameters, $k_X$ and $\delta_{\rm GS}$, are fixed by the anomaly cancellation condition (\ref{anom3})).

As discussed in ref.~\cite{Achucarro:2006zf}, the value of $W_0$ sets the overall scale of the F-potential. On the other hand, from eq.~(\ref{VDIIB}), the size of $V_{\rm D}$ is always ${\cal O}(N^{-1} T_{\rm R}^{-3})$ in Planck units. Therefore,
too large values of $W_0$ result in a too large (and negative) $V_{\rm F}$, then the uplifting by $V_{\rm D}$ is not efficient enough to promote the minimum from negative to positive. Conversely, too small values of $W_0$ would imply that $V_{\rm D}$ dominates too much and we lose the minimum; this explains the allowed range of $W_0$ values shown in fig.~\ref{W0} for the example at hand. Notice, also from that figure, that the natural scale for the potential is about five or six orders of magnitude below the Planck scale. Consequently, to obtain a Minkowski vacuum (or de Sitter with a small cosmological constant consistent with observation) requires the tuning of $W_0$, as usual.
\begin{figure}[ht]
\centering						
\includegraphics[angle=\degrees,width=12cm]{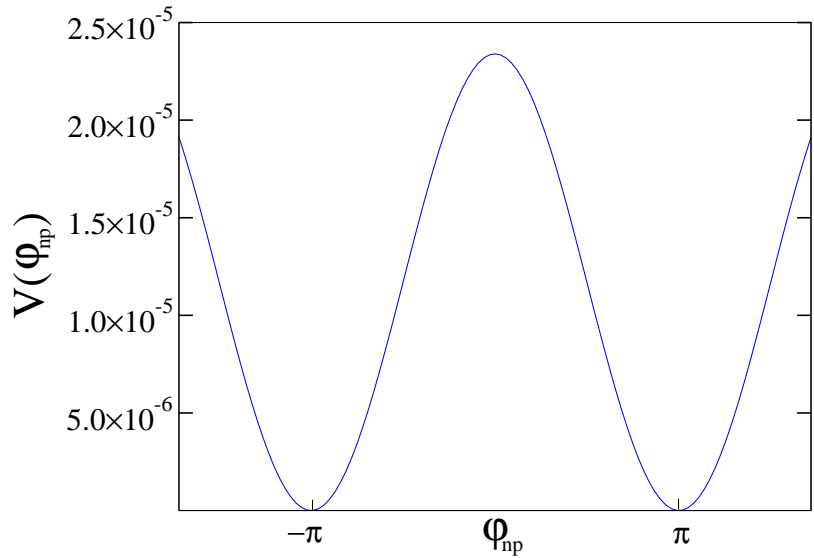}
\caption{{Periodic structure of the potential along the relative phase $\varphi_{\rm np}$. The model is the same as in the previous figure and $(T_{\rm R},\rho_{\rm M}^2)$ have been fixed to their 
at the minimum.}}
\label{pot_fase}
\end{figure}
Concerning the dependence of the potential on the other fields involved in the problem, $V$ increases monotonically  with
$\rho_{\rm M}$ as this departs from its value at the minimum, and has a periodic behaviour along the axionic direction defined by the relative phase $\varphi_{\rm np}$ given in eq.~(\ref{phase}) (the other independent phase does not appear in the potential, as already mentioned). The latter behaviour is represented in fig.~\ref{pot_fase}, which shows (for the same example as in fig.~\ref{W0}) the scalar potential as a function of 
$\varphi_{\rm np}$, with $T_{\rm R},\rho_{\rm M}^2$ fixed to their values at the minimum.

\section{Chances and problems to implement inflation}
\label{sec:etaic}

Now we consider the problem of finding a plausible candidate to inflaton in these scenarios. For that matter it is convenient to recall the most ubiquitous obstacles that one finds to implement inflation in 
concrete models.

\subsection{The $\eta$ and initial-condition problems}

In supersymmetric theories the inflaton, $\phi$, is generically one (real) component of
some complex scalar field, $\Phi$.
Successful inflation requires that the slow-roll parameters defined (in Planck units) as 
\be
\epsilon\equiv {1\over 2}\left[\frac{V'}{V}\right]^2,\;\;\;\;\;\;\;\;
\eta \equiv \frac{V''}{V} \;\;,
\label{eta}
\ee
(the primes denote differentiation with respect to $\phi$) must be $\ll 1$. On the other hand, it is well known (for a good review see ref.~\cite{Copeland:1994vg}) that Supergravity theories generically present the so-called `$\eta$ problem', which refers to the fact that $\eta$ is naturally of order one due to the presence of the ${\rm e}^K$ factor in eq.~(\ref{pot}). 

Take, for example, the simplest form of the K\"ahler potential, i.e. $K=|\Phi|^2$, which gives rise to canonical kinetic terms for $\Phi$. If we write eq.~(\ref{pot}) as 
\be
V = {\rm e}^{K}\; \tilde{V} \;\;,
\label{poteta}
\ee
and expand the exponential to first order, we see that
\be
V \sim (1+|\Phi|^2+\ldots)\;\tilde{V} \;\;.
\label{Vexp}
\ee
In other words, $\Phi$, and thus $\phi$, acquires a mass term of order $\tilde{V}$ (in Planck units) and the corresponding contribution to $\eta$ in (\ref{eta}) is of order one, which totally disfavours slow roll.

Some mechanisms have been proposed to avoid or alleviate this $\eta$ problem. In particular it is obvious that if $\phi$ does not enter the K\"ahler potential, $K$, it will not pick up a mass term coming from the expansion of the ${\rm e}^K$ term. However, this is neither sufficient nor necessary to avoid the $\eta$ problem, since one also has to take into account the possible dependence on $\Phi$ of $\tilde{V}$ in eq.~(\ref{poteta}). Schematically,
\be 
\label{fullexp}
V \sim  (1+|\Phi|^2+\ldots) \left[\tilde{V}\right]_{\Phi=0} \ + \ {\rm e}^K \left[\frac{\partial^2 \tilde{V}}{\partial \Phi \partial \bar{\Phi}}\right]_{\Phi=0}|\Phi|^2 \
+ \ \cdots \;\;.
\ee
Consequently, what should be required is that the various $|\Phi|^2$ terms involved in 
(\ref{fullexp}) cancel among themselves.

Beside the $\eta$ problem, we must worry about the initial conditions for the inflaton. In other words, it is desirable that we do not need to invoke tuning or unreasonable assumptions of any kind about the initial value of $\phi$ required to have a successful subsequent period of inflation. Note that this is typically a problem of naturalness. To that respect, an attractive possibility is that of eternal topological inflation~\cite{Vilenkin:1994pv,Linde:1994hy,Linde:1994wt}, which is realised in models where the inflaton potential has a saddle point between different, degenerate vacua. This gives rise to domain walls forming and, in the presence of an expanding universe, the false vacuum inside the walls serves as a site of inflation. This is both topological and eternal because, even though the field will be driven towards one of the minima, the core of the wall grows exponentially with time, providing us with a very generic initial state.
Whether topological inflation actually takes place or not for a given potential is a delicate 
question which has been studied for only a few examples ~\cite{Sakai:1995nh}. Generally speaking,
a sufficiently flat potential (satisfying the slow-roll conditions) with a large VEV looks to be necessary, according to these previous results.

\subsection{Candidates to inflaton}

Figs~\ref{W0}, \ref{pot_fase} suggest that both $T_{\rm R}$ and $\varphi_{\rm np}$ could be inflaton candidates. The structure of the potential along both directions shows a maximum (storing large potential energy), which might be the starting point of topological inflation. However things are not that smooth. The $T_{\rm R}$ field has an obvious $\eta$ problem since it appears explicitly in the K\"ahler potential, $K$, given in
eq.~(\ref{KIIB}) (and the same holds for the matter condensate, $\rho_{\rm M}$). The chances of $\varphi_{\rm np}$ could be better in principle since it does not appear in $K$. Actually, using $T_{\rm I}$ (which is one of the components of $\varphi_{\rm np}$, see eq.(\ref{phase})) as the inflaton was the basic
strategy followed in refs~\cite{Blanco-Pillado:2004ns, Blanco-Pillado:2006he}
(where they considered a setup with no matter fields). However, as discussed above,
this is not a guarantee to avoid the $\eta$ problem due to the presence of other dangerous contributions to $\eta$ [see eqs.~(\ref{poteta}, \ref{fullexp})]. 

Let us analyze this possibility in more depth. For the sake of the simplicity, let us 
ignore for the moment the matter fields, writing $W_{\rm np}= A e^{-aT}$. Then the $T_{\rm I}-$dependent terms in $V$ are proportional to $\cos({-a T_{\rm I}})$, having a similar size to other terms and, thus, to the whole $V$. Noting that the canonically normalized field is $\hat T_{\rm I}= (\sqrt{6}/2 T_{\rm R}) T_{\rm I}$, it is straightforward, from (\ref{fullexp}), to see that $\eta={\cal O}(1)\times (a T_{\rm R})^2$. Since, usually, $a T_{\rm R}> {\cal O}(1)$ to provide the required
suppression for the potential, then $\eta$ is naturally larger than ${\cal O}(1)$, which prevents $\hat T_{\rm I}$ from inflating. Actually, the $\eta$ problem was also found in refs.~\cite{Blanco-Pillado:2004ns, Blanco-Pillado:2006he}, and was solved by tuning the parameters of the model appropriately~\footnote{The setup in these references has two different (racetrack) exponentials in $W$, which makes the structure of $V$ more involved, but the previous schematic argument still applies.}.

In our case, we can be more precise since
the size of $W_{\rm np}$ is greatly constrained from the above-mentioned fact that
the size of $V_{\rm F}$ must be of the same order as $V_{\rm D}$, and the latter is basically 
fixed~\footnote{In refs.~\cite{Blanco-Pillado:2004ns, Blanco-Pillado:2006he} the uplifting of the potential was provided by an explicitly non-supersymmetric term $\delta V = E/T_{\rm R}^2$ which, in the spirit of KKLT, could  arise from anti D3-branes. Unlike our uplifting $V_{\rm D}$ potential,  the size of $E$ is not constrained (or it is uncertain), so it represents an extra degree of freedom that can be tuned.
Besides, the exponents of $W_{\rm np}$ were taken as continuously varying quantities [contrary to our case, where they go as $(N-1)^{-1}$ in eq.(\ref{Wnp})], which allowed their tuning.
}. More precisely, from eqs.~(\ref{VDIIB}, \ref{anom3})  [see also eq.~(\ref{vd2}) in the Appendix] the size of $V_{\rm D}$ is
\be 
V_{\rm D} \sim {27\over 128\pi N k_N^3
T_R^3}{(q+\bar q)^3\over
(q^3+\bar q^3)}\simeq {{\cal O}(1)\over 15 N k_N^3T_R^3}
\;.
\label{VDIIB2}
\ee
Now, to estimate $W_{\rm np}$ we can use the condition $V_{\rm F}\sim V_{\rm D}$ (necessary
for a successful uplifting). Actually, since $V_{\rm F}$ is a sum of terms, none of these
should exceed $V_{\rm D}$ unless there are unlikely delicate cancellations
between them. So we can concentrate, e.g., on the term $e^K K^{T\bar T} |W_T|^2$,
which depends on $W_{\rm np}$ in a nitid way. From eq.~(\ref{Wnp}), the value 
of $W_{\rm np}$ is proportional
to $\rho_{\rm M}^{-2/(N-1)}$. Usually $\rho_{\rm M}^2$ is small, but for 
$N={\cal O}(10)$ this factor is ${\cal O}(1)$ (or maybe larger). Then the condition
$e^K K^{T\bar T} |W_T|^2 \llsim V_{\rm D}$ translates into
\be
e^{-8\pi k_N T_{\rm R}/(N-1)}\llsim {\cal O}(10^{-3}) \times {1\over N (k_N T_{\rm R})^2} \;\;.
\label{condaprox}
\ee
For example for $N, T_{\rm R} = {\cal O}(10)$, which is a reasonable choice, the condition becomes $k_N T_{\rm R} /(N-1)\sim 1/2$ which, in turn, implies $\eta={\cal O}(20)$. This result is quite robust since, for other ranges of values of $N$ and $T_{\rm R}$, the exponent in eq.~(\ref{condaprox})
cannot change much if the balance between F and D terms is to be maintained. These results are confirmed by our numerical analysis.

Therefore, we can conclude that, in its simplest form, the scenario does not contain a suitable inflaton. This suggests to explore natural modifications to the simplest setup, which we do in the next section.

\section{Our model}
\label{sec:model}

The most obvious extensions of the model are to allow either more matter flavours (until now we have fixed $N_f=1$) or a second gaugino condensate, but in both cases the problems persist. 

Another natural modification is to add matter charged under the anomalous U(1) group and coupled to the squark condensate. Then the superpotential $W=W_0 + W_{\rm np}$, with 
$W_{\rm np}$ given by eq.~(\ref{Wnp}), can get extra terms such as
\be
W \rightarrow W + \lambda M^a X^b \;\;,
\label{extra}
\ee
where $\lambda$ is a coupling constant, and $X$ is a new singlet. This potential exhibits several degenerate vacua for $a>0$, $b \geq 2$ and non zero values of the singlet $X$, and a saddle point at $<X>=0$, which could be useful for the implementation of topological inflation. However, as already pointed out in ref.~\cite{Achucarro:2006zf}, this kind of coupling between singlet and matter condensate forces the anomalous charges of $X$ and $M$ to be such that the Fayet--Iliopoulos D-term can cancel. In consequence, uplifting does not happen and there is no Minkowski (or de Sitter with small cosmological constant) minimum towards which the inflaton can slow-roll.

Another possibility would consist of changing the K\"ahler potential for the matter condensate $M$, which has been, up to now, taken to be canonical [see eq.~(\ref{KIIB})], without introducing any extra fields. We looked at the possibility of considering a more string-motivated ansatz, namely
\be 
K=-3\log(T+\bar{T}-|M|^2) \;\;,
\;\;
\ee
and we checked that the shape and position of the extrema were similar to those obtained using eq.~(\ref{KIIB}). Therefore this modification of the original model would not turn any components of $T_{\rm R}$ or $M$ into a suitable inflaton.

Finally, a natural and simple extension is to consider just an additional neutral singlet, $\chi$, which is not coupled to the SU($N$) sector, except gravitationally ($\chi$ may be a superfield from another sector of the setup). The absence of terms in the superpotential coupling the singlet $\chi$ to the $T$ and $M$ fields implies that eq.~(\ref{nosusy}) holds and, as in the previous model, there are no supersymmetric 
vacua. Furthermore let us assume, for simplicity, that $\chi$ has canonical K\"ahler potential, $\Delta K = |\chi|^2$, and a polynomial superpotential $\Delta W (\chi)= \sum \lambda_n \chi^n$. If $\chi$ is to play the role of
the inflaton it is convenient (in order to implement a topological inflation mechanism) that the potential has degenerate vacua with different $\chi$ values. This condition is fulfilled, for example, when $\Delta W (\chi)$ possesses some discrete symmetry and the singlet takes a vacuum expectation value.
Again for the sake of simplicity, we will assume that the model has a $Z_2$ symmetry
$\chi \rightarrow -\chi$. Then, the complete superpotential (in $M_{\rm Planck}$ units) reads
\bea
 W & = & W_0+ W_{\rm np} + \Delta W (\chi^2) \nonumber  \\ 
   & = & W_0+ W_{\rm np} + \lambda_2 \chi^2  + \lambda_4 \chi^4 + \lambda_6 \chi^6 \;\;,
   \label{supsing}
\eea
with $W_{\rm np}$ given by eq.~(\ref{Wnp}). Higher other terms can be added but they get more and more irrelevant as long as $\langle \chi\rangle<M_{\rm Planck}$. In summary, the model is characterised
by the superpotential (\ref{supsing}) and the K\"ahler potential
\be
\label{KIIB2}
K = -3 {\rm log}(T+\bar{T})  + |M|^2 + |\chi|^2 \;.
\ee
The independent parameters are $W_0$ and $\lambda_i$. Besides, there are the parameters 
defining the gauge sector, though these should be around their natural values:
$N \sim {\cal O}(10)$, $k_N\sim {\cal O}(1)$ and $q, \bar q\sim {\cal O}(1)$.
Let us examine next the physics resulting for this
simple extension of the initial setup, as far as inflation is concerned.

\subsection{The potential}

We will first study the structure of the extrema of the scalar potential, after the addition of the $\chi$ singlet. Notice that, due to the $Z_2$ symmetry, any extremum in the modulus-condensate sector  is still an extremum of the enlarged potential for $\chi=0$.  The stability of these extrema will depend on the parameter $\lambda_2$. In particular, for sufficiently small $\lambda_2$ the extrema become saddle points.  This can be understood by fixing $\lambda_2=0$. Then, from eq.~(\ref{pot}),
\be
\label{DeltaV}
\Delta V = \left[ V_{\rm F} + e^K |W|^2\right]_{\chi=0} |\chi|^2\ +\ {\cal O}(|\chi|^4) \;\;.
\ee
The value of $V_{\rm F}$ at $\chi=0$ is the same as in the original dS or Minkowski vacuum; thus, as explained in sect.~2, $V_{\rm F} < 0$. As a matter of fact, in these scenarios the
$V_{\rm F}$ term in eq.~(\ref{DeltaV}) dominates  
over the second one within the brackets, giving rise to an instability in the singlet direction. It is clear that the slope of this instability is reduced once we switch on the mass term ($\lambda_2\neq 0$) in eq.~(\ref{supsing}).

Concerning the implementation of inflation, we can in principle use this saddle point as the 
source of eternal
topological inflation. Then the original vacuum must be a de Sitter (not Minkowski) vacuum. 

In this setup, playing with $\lambda_2$, a value for the $\eta$ parameter consistent with
slow-roll and observational data for the spectral index, $n_s\simeq 1+2\eta\simeq 0.95$, can be easily obtained although, admittedly, this represents a certain tuning of $\lambda_2$. Besides the initial saddle point, we need of course an actual minimum of the potential corresponding to the physical "quasi-Minkowski" vacuum, towards which the inflaton could roll. A minimum for large enough $|\chi|$ is actually quite easy to generate due to the quadratic (and higher order) terms in eq.~(\ref{supsing}). Then, the sign and size of $\lambda_4$, $\lambda_6$ can be chosen so that the minimum does correspond to a 
quasi-Minkowski vacuum, as desired. Of course this represents a new tuning and, in this case, a very severe one, though this is nothing but the usual fine-tuning  to adjust the phenomenological cosmological constant. Finally, the values of $W_0$ and $\lambda_i$
must be tuned in order to reproduce the size of the observed power-spectrum, $P(k)\sim
10^{-10}$ (more details will be given in the next subsection). In the analyzed examples this requires $V$ at the saddle point to be  $\sim {\cal O}(10^{-16})$.

Let us present now one example where all these conditions are fulfilled. The modulus-condensate sector is the same that the one described in section~\ref{sec:rev} . The other parameters defining the superpotential are
$
W_0 = 0.4204$, $\lambda_2 = -0.215$, $\lambda_4 = -0.055$, $\lambda_6 = -0.009
$.
The potential derived from this model has:
\begin{enumerate}
\item a saddle point at        $T_{\rm R}=18.6407$,  $
\rho_{\rm M} = 0.03551  $, $\chi =0 \ \ ,$ 
\item a quasi-Minkowski minimum at   $T_{\rm R}= 18.6554 $,  $\rho_{\rm M}= 0.03549$, $\chi = 0.0908
\ \ .$
\end{enumerate}
Notice that the $T_{\rm R}$ and $M$ fields
vary less than one per mil from one extremum to the other. Therefore we expect
that the singlet will be the main inflaton component.

The potential depends on six variables, $T_{\rm R}, T_{\rm I},\rho_{\rm M}, \alpha_{\rm M}, \chi_{\rm R}, \chi_{\rm I}$ and it is, therefore, impossible to draw a picture to illustrate how these extrema are connected. However, some simplifications are possible. $T_{\rm I}$ and 
$\alpha_{\rm M}$ only enter the problem through the $\varphi_{\rm np}$
combination, i.e. the phase of the non-perturbative superpotential  given in eq.~(\ref{phase}). Actually, for positive $W_0$, the potential is minimised at $\varphi_{\rm np}=\pi$, as in the case without singlet. (That means that $-W_0$ and $W_{\rm np}$ are aligned, partially canceling each other.)  
Analogously, the potential is minimized by taking $\chi_{\rm I}\rightarrow 0$
(for more details see the Appendix). 
Therefore, we can integrate out all phases/imaginary parts, so that the potential depends on 
($T_{\rm R}$, $\rho_{\rm M}$, $\chi_{\rm R}$). The explicit form is given in the Appendix. We just need one more step to illustrate the potential along which the inflaton ($\equiv$ singlet) evolves.  If the inflaton scale and effective mass along the slow-roll are much smaller than the $T_{\rm R}$ and $\rho_{\rm M}$ masses we can also integrate out the latter fields. In fact, since
the $T_{\rm R}$ and $\rho_{\rm M}$ masses are not far from $M_{\rm Planck}$,  this has to be the case if 
$|\eta|\ll 1$, as required for slow-roll to happen~\footnote{Incidentally, this also means
that we can focus on the scalar power spectrum, since the isocurvature fluctuations are negligible, given the hierarchy of scales between the inflation and the other fields in the system.}.

\begin{figure}[ht]
\centering						
\includegraphics[angle=\degrees, width=12cm]{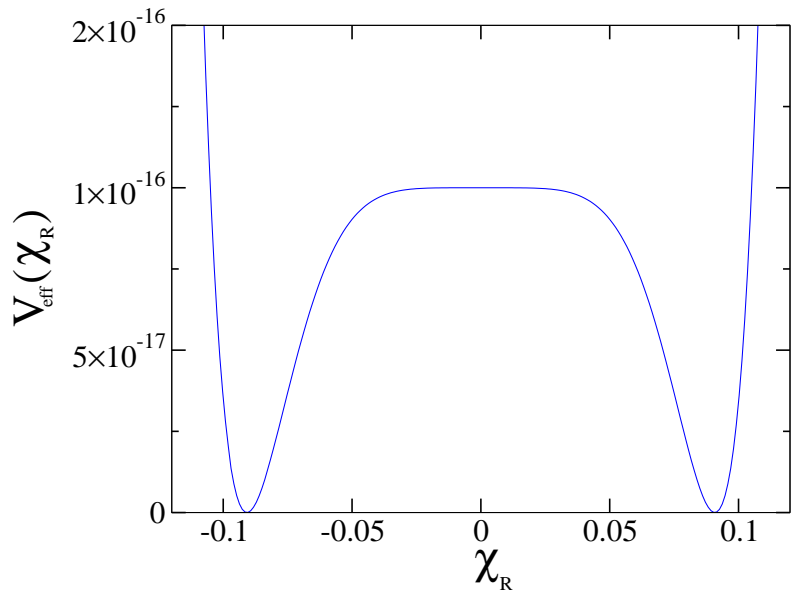}
\caption{{Effective potential as a function of $\chi_{\rm R}$}}
\label{potReChi}
\end{figure}

Consequently, we can define $ T_{\rm R}(\chi_{\rm R}), \rho_{\rm M}(\chi_{\rm R})$ through
\bea
\partial_{T_{\rm R}} V(T_{\rm R},\rho_{\rm M},\chi_{\rm R})   & = & 0  \;\;, \nonumber\\
\label{intout} &&\\
\partial_{\rho_{\rm M}} V(T_{\rm R},\rho_{\rm M},\chi_{\rm R}) & = & 0  \;\;, \nonumber
\eea
and finally write
\be
V_{\rm eff} (\chi_{\rm R}) \equiv V(T_{\rm R}(\chi_{\rm R}), 
\rho_{\rm M}(\chi_{\rm R}), \chi_{\rm R})
 \;\;.
\ee
The resulting potential is shown in fig.~\ref{potReChi}. It is of order ${\cal O} (10^{-16})$ in $M_{\rm Planck}$ units, which means that it involves mass scales of order ${\cal O} (10^{-4})$ \footnote{This is a couple of orders of magnitude below the threshold 
to produce gravity waves observable in future experiments \cite{obserwaves}.}.
The potential is almost flat in a wide region between $\chi=0$ and the minimum.
Given the relatively large VEV of $\chi_{\rm R}$, we consider this potential suitable for
implementing topological inflation.

\subsection{The equations}

After analysing the potential we are ready to write and solve the equations describing the evolution of the matter and gravitational fields during the rolling from the saddle to the minimum.
We will work in Planck units, $M_{\rm Planck} = 1$.

It is convenient to use here real (rather than complex) matter fields, i.e. 
$\{\Phi_i\}_{i=1,..., N}\rightarrow \{\phi_i\}_{i=1,..., 2N}$, where $N$ denotes the
number of complex scalar fields. Then the SUGRA Lagrangian  can be written as
\be
\label{Lmatter}
{|g|}^{-1/2}{\cal L}_{\rm matter} =  {K}_{ij} \, g^{\mu \nu} \partial_\mu \Phi^i \partial_\nu \bar \Phi^{j} - V  
=  \frac{1}{2}{\cal G}_{ij} \, g^{\mu \nu} \partial_\mu \phi^i \partial_\nu \phi^{j} - V \ ,
\ee
where $V$ is given by eq.~(\ref{VVFVD}). The relation between $K_{ij}$ and 
${\cal G}_{ij}$ can be straightforwardly calculated.

On the other hand, we parametrise the space-time metric, $g_{\mu \nu}$, as
\be
ds^2 = dt^2 - a(t)^2 dx_i dx^i \;\;.
\ee
Then, the field equations (for constant fields in space) deduced from the matter and the gravity Lagrangians are     
\be
\label{hom1}
\ddot{\phi}^{i}\,+3\,H \,\dot{\phi}^{i}
+\,\Gamma^{i}_{jk}\, \dot{\phi}^{j}\,\dot{\phi}^{k}\,+{\cal G}^{ij}\,\frac{\partial V}{\partial\phi^{j}}=0 \;\;,
\ee
subject to the constraint
\be
\label{hom2}
H^2 = \left(\frac{\dot{a}}{a}\right)^2=\frac{1}{3}\left[ \frac{1}{2}\,{\cal G}_{ij}\,\dot{\phi}^{i}\dot{\phi}^{j}+V\right] \;\;,
\ee
where dots denote time derivatives and, as usual, $\Gamma_{jk}^i$ are the Christoffel symbols derived from the ${\cal G}_{ij}$ metric 
\be
\Gamma^{i}_{jk}=\frac{1}{2}\,{\cal G}^{im}\left[ \frac{\partial {\cal G}_{mk}}{\partial \phi^{j}}+\frac{\partial {\cal G}_{jm}}{\partial \phi^{k}}-\frac{\partial {\cal G}_{jk}}{\partial \phi^{m}} \right] \;\;,
\ee
with ${\cal G}^{ij}{\cal G}_{jk}=\delta^i_k$. In order to determine the evolution
of the matter fields, we perform a change of independent variable from time to number of e-folds, 
$N_{e}$
\be a(t)=e^{N_{e}(t)}\,\,\,,\,\,\,\,\,\, H=\frac{dN_{e}(t)}{dt} \;\;.
\ee
Then, from (\ref{hom1}, \ref{hom2}) we can write the evolution equations for the
matter fields as~\footnote{In the slow-roll approximation eq.~(\ref{evoleq}) 
simplifies to
$\;\;\;
\phi'^i + {\cal G}^{ij} \frac{1}{V} \frac{\partial V}{\partial \phi^j} = 0 \;\;.$}
\be
\label{evoleq}
\phi^{i}{''}+\left[1-\frac{1}{6}{\cal G}_{jk}\phi^{j}{'}\phi^{k}{'}\right]\left[3\phi^{i}{'}+3{\cal G}^{ij}\frac{1}{V}\left(\frac{\partial V}{\partial\phi^{j}}\right)\right]+\Gamma^{i}_{jk}\, \phi^{j}{'}\phi^{k}{'}=0 \;\;,
\ee
where prime means derivative respect to $N_{e}$. Note that in this way the scale factor is no longer present in the evolution equations.

In our case we have six real component fields, $\{\phi_i\}\equiv\{T_{\rm R}, T_{\rm I}, \rho_{\rm M}, \alpha_{\rm M}, \chi_{\rm R}, \chi_{\rm I} \}$. Although we have performed
all the numerical computations with the complete set of $\phi$ fields, 
it is possible to decouple $\{T_{\rm I}, \alpha_{\rm M}, \chi_{\rm I} \}$ since,
as argued in the previous subsection, they minimize the potential at well-defined
values independent of the value of the other fields. So they rapidly fall into their
minimizing values and play no role in the evolution of the other fields. 
Then, the matter Lagrangian (\ref{Lmatter})
for the three relevant fields, $\{T_{\rm R}, \rho_{\rm M}, \chi_{\rm R} \}$,
has the form
\be
|g|^{-1/2}{\cal L}_{\rm matter} = \left[\frac{3}{4T_{\rm R}^2}\,\partial_\mu T_{\rm R} \, \partial^\mu T_{\rm R}+\partial_\mu 
\rho_{\rm M} \, \partial^\mu 
\rho_{\rm M}+\partial_\mu \chi_{\rm R} \, \partial^\mu \chi_{\rm R}- V \right] \;\;,
\label{lag3}
\ee
where $V$ is explicitly given in the Appendix. The evolution equations
(\ref{evoleq}) applied to these fields read
\be
\chi_{\rm R}''+\left[1-\frac{1}{3}\,{\chi_{\rm R}'}^{2}-\frac{1}{4 T_{\rm R}^2}\,{T_{\rm R}'}^2-\frac{1}{3}\,{
\rho_{\rm M}'}^2\right]  \left[3\,\chi_{\rm R}'+\frac{3}{2V}\,\left( \frac{\partial V}{\partial \chi_{\rm R}}\right) \right]=0 \;\;,
\ee
\be
T_{\rm R}''+\left[1-\frac{1}{3}\,{\chi_{\rm R}'}^{2}-\frac{1}{4 T_{\rm R}^2}\,{T_{\rm R}'}^2-\frac{1}{3}\,{
\rho_{\rm M}'}^2\right]  \left[3\,T_{\rm R}'+\frac{2T_{\rm R}^2}{V}\,\left( \frac{\partial V}{\partial T_{\rm R}}\right) \right]=\frac{{T_{\rm R}'}^2}{T_{\rm R}} \;\;,
\ee
\be
\rho_{\rm M}''+\left[1-\frac{1}{3}\,{\chi_{\rm R}'}^{2}-\frac{1}{4T_{\rm R}^2}\,{T_{\rm R}'}^2-\frac{1}{3}\,{
\rho_{\rm M}'}^2\right]  \left[3\,
\rho_{\rm M}'+\frac{3}{2V}\,\left( \frac{\partial V}{\partial 
\rho_{\rm M}}\right) \right]=0 \;\;.
\ee
\begin{figure}[ht]
\centering						
\includegraphics[angle=\degrees, width=12cm]{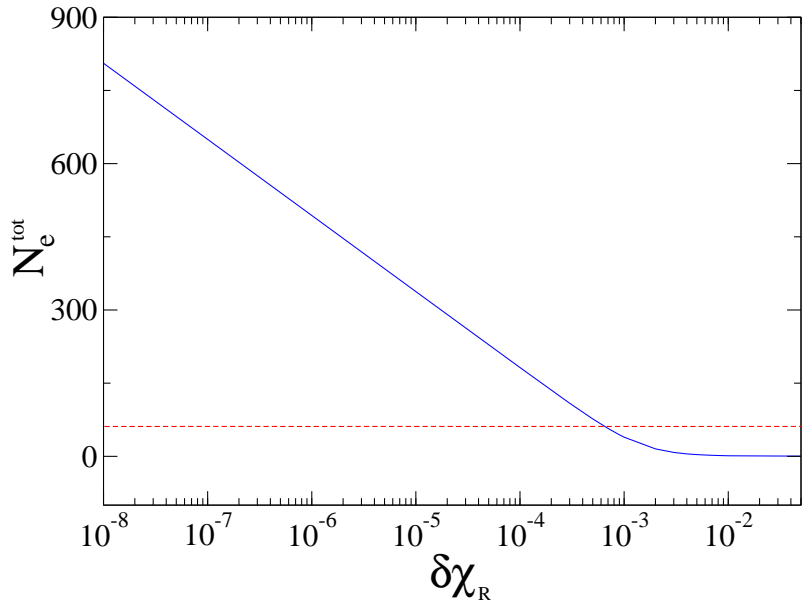}
\caption{{ Plot of the total number of e-folds of inflation, $N_e^{\rm tot}$, as a function of the initial condition for the inflaton quoted as the shift,  $\delta \chi_R$, with respect to its value at the saddle point given by $\chi=0$. The dotted line indicates the 60 e-folds needed to make inflation successful. }}
\label{duracion_inflacion}
\end{figure}
As argued in the previous section, the inflaton corresponds essentially to the 
$\chi_{\rm R}$ field. The total number of e-folds, $N_e^{\rm tot}$, depends on the initial condition 
for $\chi_{\rm R}$. If, initially, $\chi=0$, then $N_e^{\rm tot}\rightarrow \infty$;
otherwise $N_e^{\rm tot}$ depends on the initial shift, $\delta \chi_{\rm R}$. Fig.~\ref{duracion_inflacion}.
shows the dependence of $N_e^{\rm tot}$ on $\delta \chi_{\rm R}$
using the values of the parameters of the model given in the previous subsection.
Note that $N_e^{\rm tot}\geq 50-60$, as
phenomenologically required corresponds to $\left. \delta\chi_{\rm R}\right|_{\rm initial}\llsim 10^{-3}$.
Recall here that, since
we are using the saddle point at $\chi=0$ as the origin of topological inflation,
this means that all the initial conditions are realized in practice (they correspond
to different spatial points inside the associated domain wall). The final stages
of inflation are the same for all of them, the only difference being the total
number of e-folds before the end of inflation. In consequence, any region 
of the domain wall with $\left. \delta\chi_{\rm R}\right|_{\rm initial}\llsim 10^{-3}$
gives appropriate inflation, with no tuning of initial conditions.

\begin{figure}[ht]
\centering						
\includegraphics[angle=\degrees, width=12cm]{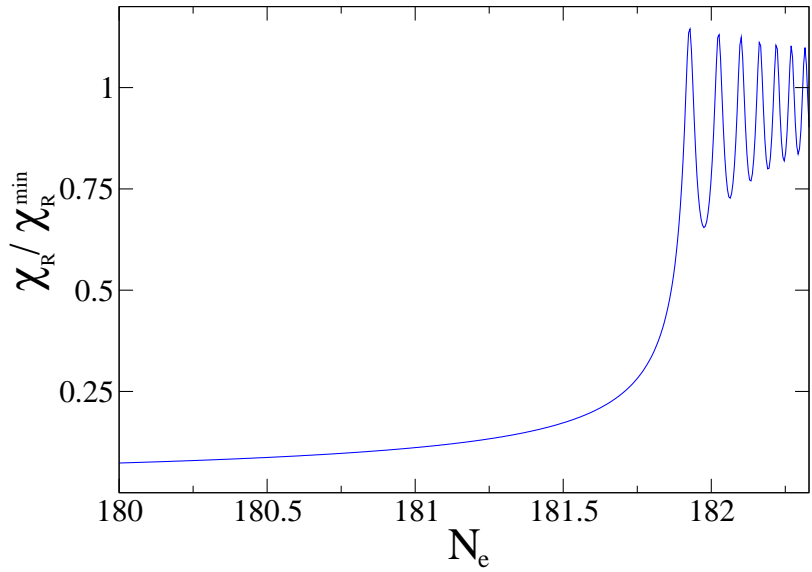}
\caption{{Cosmological evolution of the singlet $\chi_{\rm R}$, normalised to its minimum value, as a function of the number of e-folds, $N_e$.}}
\label{singlet_1}
\end{figure}

\begin{figure}[ht]
\centering						
\includegraphics[angle=\degrees, width=12cm]{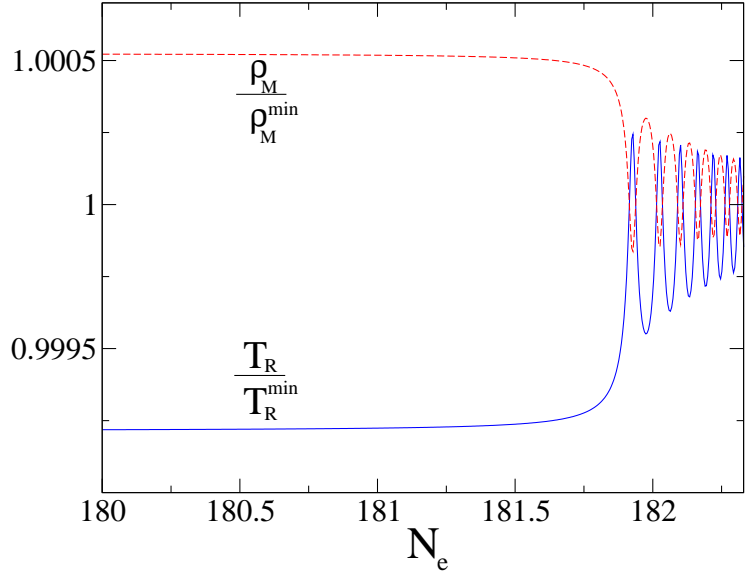}
\caption{{Cosmological evolution of $T_{\rm R}$ and $
\rho_{\rm M}$, normalised to their minimum values, as a function of $N_e$.}}
\label{tm}
\end{figure}

In order to show the evolution profiles for the singlet ($\chi_{\rm R}$),
modulus ($T_{\rm R}$) and condensate ($\rho_{\rm M}$) we have taken 
$\left. \delta\chi_{\rm R}\right|_{\rm initial}=10^{-4}$, which corresponds to
$N_e^{\rm tot}\simeq 180$, but we insist that the results are the same for any other 
initial condition (provided $N_e^{\rm tot}>60$) since the last 60
e-folds take place in the same way. The corresponding profiles are shown
in figs~\ref{singlet_1},\ref{tm}.

\begin{figure}[ht]
\centering						
\includegraphics[angle=\degrees, width=12cm]{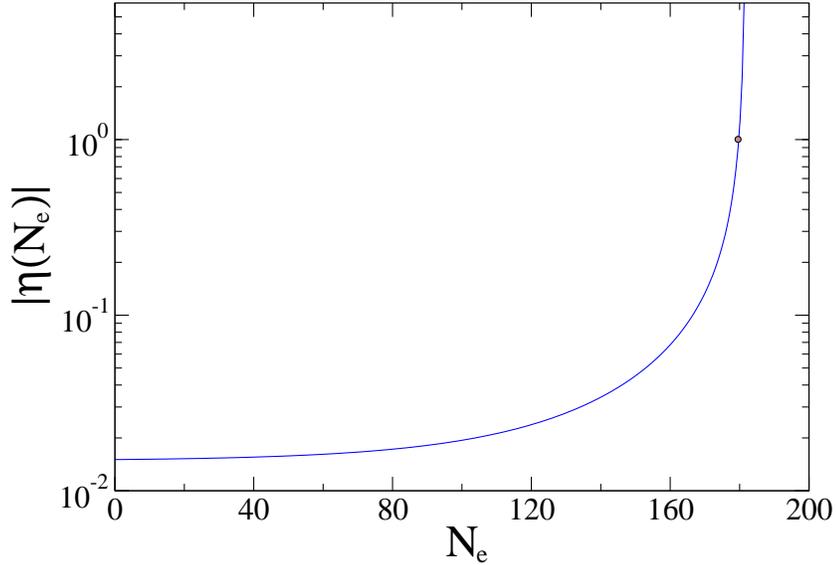}
\caption{{Absolute value of  $\eta$ as a function of $N_e$. The dot indicates $\eta=1$.}}
\label{eta_1}
\end{figure}

\begin{figure}[ht]
\centering						
\includegraphics[angle=\degrees, width=7.5cm]{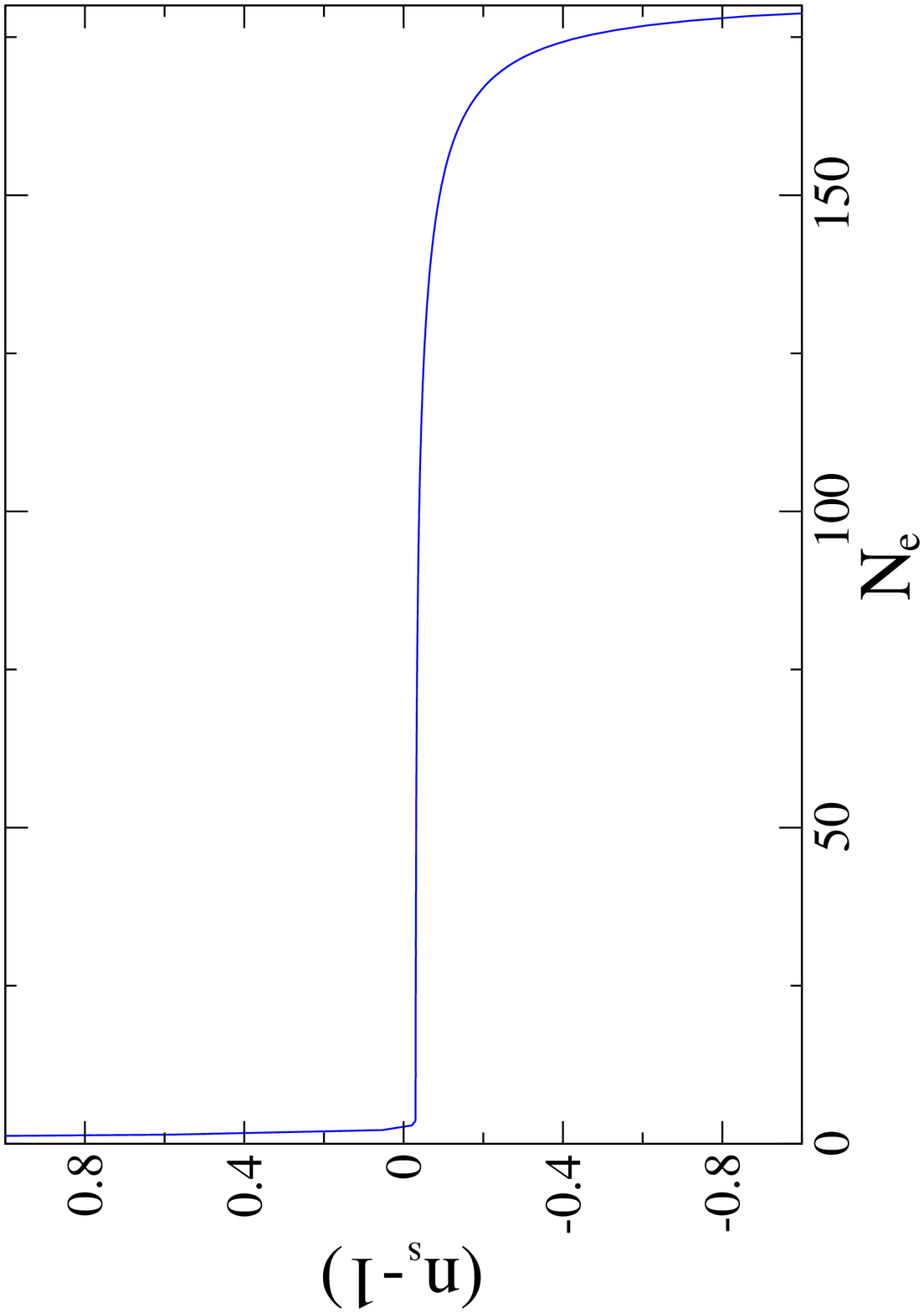}
\includegraphics[angle=\degrees, width=7.5cm]{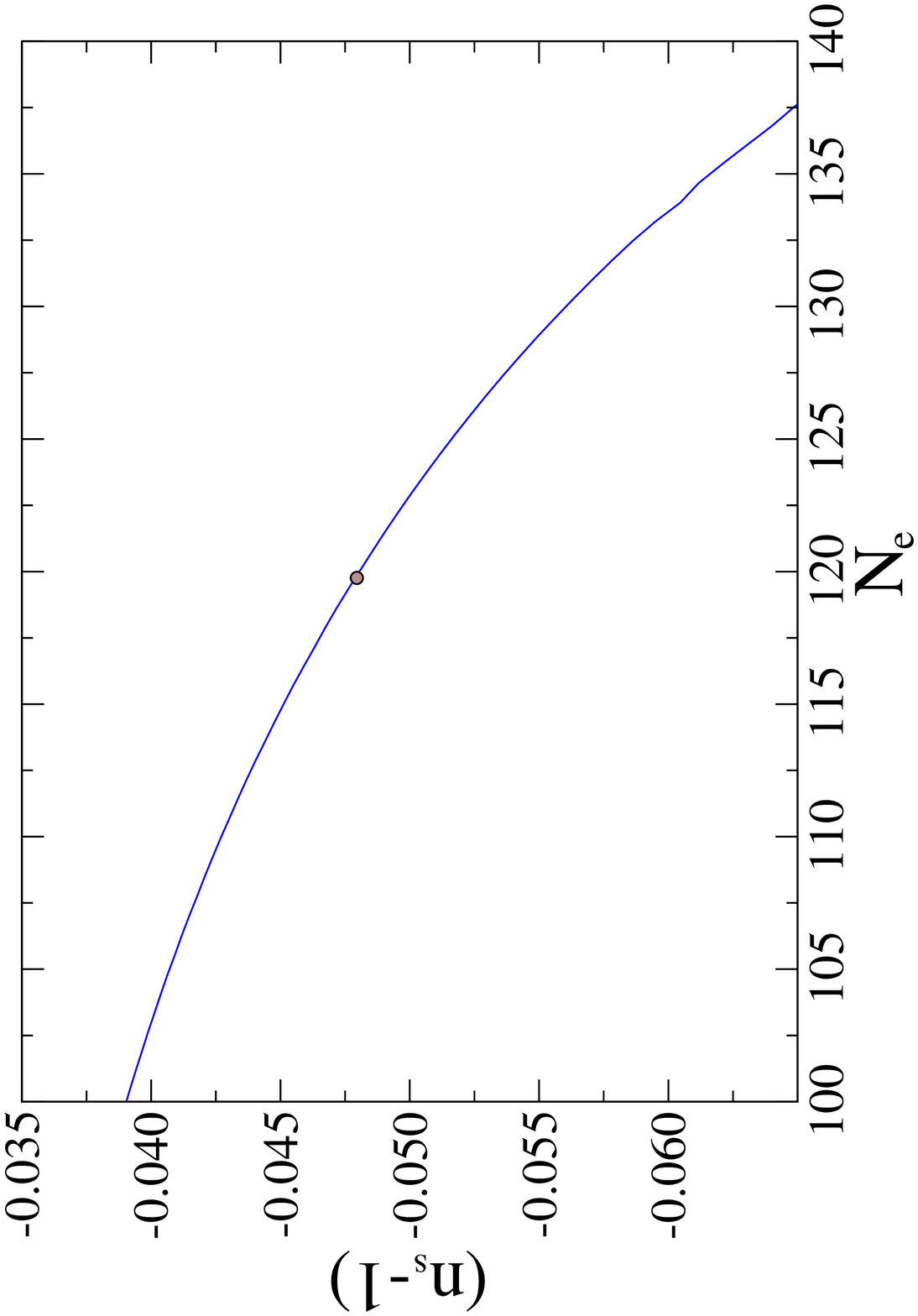}
\caption{{ Left: $n_s$ as a function of $N_e$ for the whole range of the evolution of the inflaton. Right: zoom of the region where the value of $n_s$ is provided by WMAP.}}
\label{ns}
\end{figure}

At $N_e\sim180$ the fields start to oscillate around their minimum values, signaling the end of inflation. 
To determine the end of inflation more precisely we need to evaluate the slow roll parameters $\epsilon$
\be
\epsilon = \frac{1}{2 V^2} {\cal G}^{ij}  \partial_i V \partial_j V 
         = \frac{1}{2 V^2} {\cal G}^{ij}  \frac{\partial V}{\partial \phi^i} \frac{\partial V}{\partial \phi^j} \;,
\ee
and $\eta$, defined as the most negative eigenvalue of the matrix~\cite{Blanco-Pillado:2006he}
\be
\eta^i_j = \frac{1}{V}   {\cal G}^{ik} \left(  \partial_k \partial_j V - \Gamma^l_{kj} \partial_l V \right) \;\;.
\ee
The $\epsilon$ parameter remains ${\cal O}(10^{-9})$ along the evolution.
The behaviour of $\eta$ as a function of $N_{e}$ is shown in fig.~\ref{eta_1}, from which we infer that inflation lasts until $N_e\sim 180$.
The spectral index is defined to be
\be
\label{nscalar}
n_s = 1+ \frac{d \log(P(k))}{d \log(k)} \;\;, 
\ee
where $P(k)$ is the power spectrum of scalar density perturbations~\footnote{In the slow roll approximation, $\frac{1}{2}{\cal G}_{ij}{\phi'}^i{\phi'}^j=\epsilon$, so we could compute the power spectrum as  $P(k)=\frac{1}{150\pi^2}\frac{V}{\epsilon-\frac{1}{3}\epsilon^2}\simeq \frac{1}{150\pi^2}\frac{V}{\epsilon}$. Similarly, $n_s\simeq 1+2\eta-6\epsilon+\cdots$ However, we have done the calculation without any approximation.} \cite{Blanco-Pillado:2006he}
\be
P(k) = \frac{1}{50\pi^2}\frac {H^4}{{\cal L}_{kin}}=\frac{1}{150\pi^2}\frac {V}{\left(\frac{1}{2}{\cal G}_{ij}{\phi'}^i{\phi'}^j \right) - \frac{1}{3}\left(\frac{1}{2}{\cal G}_{ij}{\phi'}^i{\phi'}^j\right)^2}\;\;.
\ee
The spectral index is shown, as $n_s-1$, in fig.~\ref{ns}. Recall that the window of  allowed values for this parameter coming from the WMAP data corresponds to it being evaluated $\sim 60$ e-folds before the end of inflation.
In our case this would mean $N_e\sim 120$, which is the region shown in detail on the right hand side of fig.~\ref{ns}. The spectral index (\ref{nscalar}) was calculated using 
\be 
n_s \simeq 1+ \frac{d \log(P_{k}(N_{e}))}{d N_{e}} \;\;,
\ee
since $d\log k\cong dN_{e}$ at horizon crossing. Note that $n_s \sim 0.95$, as required by the WMAP fit  
\cite{Spergel:2006hy}. Note also that $ d n_s / d\log k \ll 1$, which is consistent with \cite{Spergel:2006hy}.

We have arranged the parameters so that the slope and normalization of $P(k)$ (namely,
$P(k)=4\times10^{-10}$) around $60$ e-folds before the end of inflation are satisfied at the same time. 

Let us now compare our approach and results with some recent proposals which touch upon similar models. In ref.~\cite{Holman:2006tm} the issue of decoupling moduli which are apparently irrelevant for inflation in the context of Type IIB string theory compactified on a Calabi-Yau orientifold was addressed. 
Some of their methods apply to the large volume stabilisation proposed and developed in refs~\cite{Balasubramanian:2004uy,Balasubramanian:2005zx,Conlon:2005ki}. 
Inflation in that context has also been reviewed in refs~\cite{Conlon:2005jm,Bond:2006nc},
where the evolution of the imaginary part of the relevant modulus was considered.
They conclude that a minimum setup of three moduli is needed for inflation to work, and the stabilisation of two of those is assumed before starting the evolution of the third.
Although their claim is that no fine tuning is needed in these scenarios (as opposed to that of ref.~\cite{Blanco-Pillado:2006he}), we believe that there is an implicit tuning of the shape of the potential, encoded in the dynamics of the two moduli which are already stabilised.

As discussed above, the need of fine-tuning is also necessary in the scenario presented here.
This point is sustained as well by the results of ref.~\cite{Ellis:2006ar}, where an attempt was made to incorporate chaotic/new inflation within SUGRA in a string context. The characteristics of those models are similar to ours, namely a modulus $T_{\rm R}$ and an inflaton $\Phi$,  the latter with polynomial self interactions, but the mechanism for stabilising $T_{\rm R}$ is  different. The conclusions are, however, similar to ours, namely a substantial degree of fine-tuning is needed to arrange a successful inflation.

\section{Conclusions}
 \label{sec:conclu}

In this paper we have looked for a model of slow roll inflation that could be accommodated within the uplifting setup of ref.~\cite{Achucarro:2006zf}. That is, we have considered a scenario of $N=1$, D=4 SUGRA  with gauge group SU($N$)$\times$ U(1) whose gauge coupling
is proportional to a $T$ modulus. The
dynamics is determined by the SU($N$) gaugino condensation and a flux parameter.
Moreover, the necessary uplifting of the potential is
achieved thanks to a Fayet-Iliopoulos D-term, which imposes the presence of chiral matter 
transforming under the U(1). Part of the interest of this setup comes from the fact
that most of the parameters are constrained by Supersymmetry and gauge invariance,
improving its predictive power.
The corresponding scalar potential has dS and Minkowski minima for reasonable values of both modulus and matter condensate, which are natural candidates for the onset
of inflation. However, after a careful analysis, we have concluded  that the system,
in its simplest
version (as it was formulated in \cite{Achucarro:2006zf}), does {\em not} contain a suitable inflaton. This is mainly due to the severeness of the $\eta$ problem in Supergravity theories. [We have also given a review of this problem
in connection to generic modular (i.e. $T$-driven) inflation scenarios.]

Next, we have studied the most natural and economical extension of this simple framework in order to implement successful inflation, concluding that it consists of an additional singlet, $\chi$, with zero 
U(1) charge and an ordinary polynomial superpotential. 
The resulting model is quite attractive, as it incorporates eternal topological inflation and can be arranged to comply with all the WMAP and other observational constraints. The role of the
(slow rolling) inflaton is played by (the real part of) the $\chi$ field, but the
dynamics responsible for the stabilization of the moduli (i.e.  gaugino condensation, flux and D-uplifting) is crucial for inflation to take place, in particular it is  essential to achieve the required positive energy.

We find this kind of set up completely natural, and thus appealing. Nevertheless, it must be pointed out that the above-mentioned arrangement of parameters requires a considerable fine-tuning. This is unpleasant, although, as discussed in the paper, it is a generic
fact of virtually all working examples of modular inflation. This may be an indication of a general disease in inflationary scenarios based on Supergravities coming from strings,
or just a consequence of the fact 
that inflationary model building in this context is still in its infancy.
Given the obvious interest of such context, the fact that models can actually
be built is a sign of progress and should encourage further work in the field.
Finally, one can take a more optimistic (though less demanding)
view within a landscape philosophy.
Then, the apparent fine-tuning of our ``local'' universe would be a consequence of the need of inflation for the creation of large and lasting universes with matter, where
the latter can evolve to life.

\subsection*{Acknowledgments}

We thank Mark Hindmarsh, Andrew Liddle and N. Sakai for interesting discussions. 
BdC thanks the IFT (Madrid), and JMM thanks the Particle Theory group at Sussex University  for hospitality at different stages of this project. AG acknowledges the financial support of a FPI (MEC) grant,  ref.~BES-2005-8412.  This work was partially supported by PPARC, the MEC project FPA 2004-02015, the Comunidad de Madrid project HEPHACOS, ref.~P-ESP-00346, and EU research and training network MRTN-CT-2006-035863. 

\subsection*{Appendix}

In this appendix we give the explicit form of the potential needed in eq.~(\ref{lag3}) to calculate the field evolution from the vicinity of the saddle point. We recall that the model is defined in the last 
paragraph before subsect.~4.1  [see eqs~(\ref{supsing},\ref{KIIB2})].

We write the supergravity potential as a sum of the F and D parts $V = V_{\rm F} + V_{\rm D}$. 
Since the D part is associated with the anomalous U(1) it only involves the fields, $T$ and $M$.
Its expression  was given in  eq.~(\ref{VDIIB}). Using the anomaly cancellation
condition (\ref{anom3}) we can rewrite it as
\be
V_{\rm D} = \frac{3 \pi}{8 N k_N T_{\rm R}} \frac{(q+\bar{q})^3}{q^3+\bar{q}^3} 
\left( \rho_M^2 + \frac{3}{4 \pi k_N T_{\rm R} }\right)^2\;.
\label{vd2}
\ee
The F part is given by eq.~(\ref{pot}) and depends on the K\"ahler derivatives for the 
chiral superfields,
\begin{equation}
\hspace{-1.9cm}
 \left(
       \begin{array}{c}  
                       D_T W \\[2mm] 
                      M  D_M W \\[2mm] 
                    \chi  D_\chi W 
        \end{array}
  \right)
= 
  \left(
        \begin{array}{ccccc}  
                 \frac{-3}{T+\bar{T}} - \frac{4 \pi k_N}{N-1}&  \frac{-3}{T+\bar{T}} &
                \frac{-3}{T+\bar{T}} &  \frac{-3}{T+\bar{T}} &  \frac{-3}{T+\bar{T}} \\[2mm]                  
M \bar{M} - \frac{2}{N-1} & 
M \bar{M} & 
M \bar{M} & 
M \bar{M} & 
M \bar{M} 
\\[2mm]
            \chi \bar{\chi} &   \chi \bar{\chi} &   \chi \bar{\chi} + 2 &  \chi \bar{\chi} + 4& \chi \bar{\chi}  + 6
        \end{array}
  \right)
  \left(
        \begin{array}{c}  
                        W_{np} \\ 
                        W_0 \\ 
                        W_2 \\ 
                        W_4 \\
                       W_6 
        \end{array}
  \right)
\label{app2}
\end{equation}
where $W_{2,4,6}$ stand for the monomials proportional to $\chi^{2,4,6}$ respectively
in eq.~(\ref{supsing}). Notice that 
the above matrix elements are real. As a consequence, the only relevant phases
are the relative ones among the different terms appearing in the superpotential.
Besides, $T_{\rm I}$ and  $\alpha_{\rm M}$ appear only through the phase of the 
non-perturbative term, $\varphi_{\rm np}$, see eq.(\ref{phase}). 

In our case, for the range of interest of $T_{\rm R}$, $\rho_{\rm M}$ and
$\chi_{R}$ (i.e. from the saddle point to the minimum neighbourhood)
the potential is minimal for $\varphi_{\rm np} = \pi $
(we assume real $W_0$)
and $\chi_{I}=0$.
Then if we evolve the fields from an initial configuration with all the phases at the minimal value,
they will remain constant.  We can then restrict ourselves to a reduced potential, $V_{\rm F} (T_{\rm R}, \rho_{\rm M}, \chi_{\rm R})$. In terms of these three relevant variables, we get
\bea
\hspace{-1.5cm} V_{\rm F}  \hspace{-1cm} & \hspace{-.75cm} = &   
                 \frac{1}{8 T_{\rm R}^3} e^{\rho_{\rm M}^2 + \chi_{\rm R}^2} 
                  \nonumber \\ 
&  &   
  \left\{ 
          3 \left(   A - 
                    \frac{8 \pi T_{\rm R} k_N}{3}
                    \left( \frac{2}{\rho_{\rm M}^2} \right)^{\frac{1}{{N-1}}} 
                    e^{\frac{- 4 \pi k_N}{N-1}T_{\rm R}} 
            \right)^2  
         +
            \rho_{\rm M}^2 
            \left(      A +  \left( \frac{2}{\rho_{\rm M}^2} \right)^{\frac{N}{N-1}} 
                     e^{\frac{- 4 \pi k_N}{N-1}T_{\rm R}}
            \right)^2  \right. \nonumber \\
&  &  \hspace{.2cm}
      \left. 
      + \; \chi_{\rm R}^2 \left( 
                            A + 2 \lambda_2 + 4 \lambda_4 \chi_{\rm R}^2 
                                            + 6 \lambda_6 \chi_{\rm R}^4 
                       \right)^2  - 3 A^2 \right\} \;\; ,
\label{app3}
\eea
where
\be
A = W_0 - (N-1) \left( \frac{2}{
\rho_{\rm M}^2} \right)^{\frac{1}{{N-1}}} e^{\frac{- 4 \pi k_N}{N-1}T_{\rm R}}
    + \lambda_2 \chi_{\rm R}^2   +  \lambda_4 \chi_{\rm R}^4 +  \lambda_6 \chi_{\rm R}^6 \;\; .
\ee
For the case without the singlet $\chi$, the previous eqs.~(\ref{app2}, \ref{app3})
hold, taking $\chi\rightarrow 0$.


\section*{References}
\begin{thebibliography}{100}
\bibitem{Kachru:2003aw}
S.~Kachru, R.~Kallosh, A.~Linde and S.~P.~Trivedi,
Phys.\ Rev.\ D {\bf 68}, 046005 (2003)
[arXiv:hep-th/0301240].
%
\bibitem{Spergel:2006hy}
D.~N.~Spergel {\it et al.},
 arXiv:astro-ph/0603449.
%
\bibitem{Dvali:1998pa}
  G.~R.~Dvali and S.~H.~H.~Tye,
  Phys.\ Lett.\ B {\bf 450} (1999) 72
  [arXiv:hep-ph/9812483].
 %

\bibitem{Binetruy:1986ss} 
P.~Bin\'etruy and M.~K.~Gaillard,
Phys.\ Rev.\ D {\bf 34} (1986) 3069.
\bibitem{Brustein:1992nk}
R.~Brustein and P.~J.~Steinhardt,
Phys.\ Lett.\ B {\bf 302} (1993) 196
[arXiv:hep-th/9212049].
\bibitem{Copeland:1994vg}
  E.~J.~Copeland, A.~R.~Liddle, D.~H.~Lyth, E.~D.~Stewart and D.~Wands,
  Phys.\ Rev.\ D {\bf 49}, 6410 (1994)
  [arXiv:astro-ph/9401011].

\bibitem{Banks:1995dp}
  T.~Banks, M.~Berkooz, S.~H.~Shenker, G.~W.~Moore and P.~J.~Steinhardt,
  Phys.\ Rev.\ D {\bf 52} (1995) 3548
  [arXiv:hep-th/9503114].
  
  \bibitem{Bailin:1998tu}
  D.~Bailin, G.~V.~Kraniotis and A.~Love,
  Phys.\ Lett.\ B {\bf 443} (1998) 111
  [arXiv:hep-th/9808142].

\bibitem{Douglas:2006es}
  M.~R.~Douglas and S.~Kachru,
  arXiv:hep-th/0610102.
  
  
  \bibitem{Choi:2006bh}
  K.~Choi and K.~S.~Jeong,
  JHEP {\bf 0608} (2006) 007
  [arXiv:hep-th/0605108].
  
  
  \bibitem{Burgess:2003ic}
C.~P.~Burgess, R.~Kallosh and F.~Quevedo,
JHEP {\bf 0310}, 056 (2003)
[arXiv:hep-th/0309187].


\bibitem{Achucarro:2006zf}
A.~Ach\'ucarro, B.~de Carlos, J.~A.~Casas and L.~Doplicher,
JHEP {\bf 0606}, 014 (2006)
[arXiv:hep-th/0601190].


  \bibitem{Villadoro:2005yq}
  G.~Villadoro and F.~Zwirner,
  Phys.\ Rev.\ Lett.\  {\bf 95} (2005) 231602
  [arXiv:hep-th/0508167].
  
\bibitem{Parameswaran:2006jh}
  S.~L.~Parameswaran and A.~Westphal,
  JHEP {\bf 0610} (2006) 079
  [arXiv:hep-th/0602253].
  
  \bibitem{Westphal:2006tn}
  A.~Westphal,
  arXiv:hep-th/0611332.


\bibitem{Burgess:2001fx}
  C.~P.~Burgess, M.~Majumdar, D.~Nolte, F.~Quevedo, G.~Rajesh and R.~J.~Zhang,
  JHEP {\bf 0107} (2001) 047
  [arXiv:hep-th/0105204].

\bibitem{Dvali:2001fw}
  G.~R.~Dvali, Q.~Shafi and S.~Solganik,
  arXiv:hep-th/0105203.
  
  \bibitem{Burgess:2001vr}
  C.~P.~Burgess, P.~Martineau, F.~Quevedo, G.~Rajesh and R.~J.~Zhang,
  JHEP {\bf 0203} (2002) 052
  [arXiv:hep-th/0111025].

\bibitem{Garcia-Bellido:2001ky}
  J.~Garc\'{\i}a-Bellido, R.~Rabad\'an and F.~Zamora,
  JHEP {\bf 0201} (2002) 036
  [arXiv:hep-th/0112147].
  
  \bibitem{Jones:2002cv}
  N.~T.~Jones, H.~Stoica and S.~H.~H.~Tye,
  JHEP {\bf 0207} (2002) 051
  [arXiv:hep-th/0203163].
  
  \bibitem{Gomez-Reino:2002fs}
  M.~G\'omez-Reino and I.~Zavala,
  JHEP {\bf 0209} (2002) 020
  [arXiv:hep-th/0207278].

  
  
\bibitem{Dasgupta:2002ew}
  K.~Dasgupta, C.~Herdeiro, S.~Hirano and R.~Kallosh,
  Phys.\ Rev.\ D {\bf 65} (2002) 126002
  [arXiv:hep-th/0203019].

\bibitem{Kachru:2003sx}
  S.~Kachru, R.~Kallosh, A.~Linde, J.~M.~Maldacena, L.~McAllister and S.~P.~Trivedi,
  JCAP {\bf 0310} (2003) 013
  [arXiv:hep-th/0308055].


\bibitem{Hsu:2003cy}
  J.~P.~Hsu, R.~Kallosh and S.~Prokushkin,
  JCAP {\bf 0312} (2003) 009
  [arXiv:hep-th/0311077].
  
  \bibitem{Firouzjahi:2003zy}
  H.~Firouzjahi and S.~H.~H.~Tye,
  Phys.\ Lett.\ B {\bf 584} (2004) 147
  [arXiv:hep-th/0312020].
  
  
\bibitem{Burgess:2004kv}
  C.~P.~Burgess, J.~M.~Cline, H.~Stoica and F.~Quevedo,
  JHEP {\bf 0409} (2004) 033
  [arXiv:hep-th/0403119].
  
\bibitem{Blanco-Pillado:2004ns}
  J.~J.~Blanco-Pillado {\it et al.},
  JHEP {\bf 0411} (2004) 063
  [arXiv:hep-th/0406230].


\bibitem{Lalak:2005hr}
  Z.~Lalak, G.~G.~Ross and S.~Sarkar,
  arXiv:hep-th/0503178.
  
  \bibitem{Cline:2005ty}
  J.~M.~Cline and H.~Stoica,
  Phys.\ Rev.\ D {\bf 72} (2005) 126004
  [arXiv:hep-th/0508029].
  
  \bibitem{Conlon:2005jm}
  J.~P.~Conlon and F.~Quevedo,
  JHEP {\bf 0601} (2006) 146
  [arXiv:hep-th/0509012].
  
  
  \bibitem{Blanco-Pillado:2006he}
  J.~J.~Blanco-Pillado {\it et al.},
  JHEP {\bf 0609} (2006) 002
  [arXiv:hep-th/0603129].
  

\bibitem{Kallosh:2006dv}
  R.~Kallosh and A.~Linde,
  arXiv:hep-th/0611183.
  



\bibitem{Kallosh:2007ig}
  R.~Kallosh,
  arXiv:hep-th/0702059.


%
\bibitem{Choi:2005ge}
K.~Choi, A.~Falkowski, H.~P.~Nilles and M.~Olechowski,
Nucl.\ Phys.\ B {\bf 718}, 113 (2005)
[arXiv:hep-th/0503216].
%
\bibitem{deAlwis:2005tf}
S.~P.~de Alwis,
Phys.\ Lett.\ B {\bf 626}, 223 (2005)
[arXiv:hep-th/0506266].
%
\bibitem{Haack:2006cy}
  M.~Haack, D.~Krefl, D.~Lust, A.~Van Proeyen and M.~Zagermann,
  arXiv:hep-th/0609211.
 %
\bibitem{Ibanez22}
L.~E.~Ib\'a\~nez, C.~Mu\~noz and S.~Rigolin,
Nucl.\ Phys.\ B {\bf 553} (1999) 43
[arXiv:hep-ph/9812397];
D.~Lust, S.~Reffert and S.~Stieberger,
Nucl.\ Phys.\ B {\bf 706} (2005) 3
[arXiv:hep-th/0406092];
P.~G.~C\'amara, L.~E.~Ib\'a\~nez and A.~M.~Uranga,
Nucl.\ Phys.\ B {\bf 708} (2005) 268
[arXiv:hep-th/0408036].
%
  \bibitem{Taylor:1982bp}
  T.~R.~Taylor, G.~Veneziano and S.~Yankielowicz,
  Nucl.\ Phys.\ B {\bf 218} (1983) 493.
  
  
  \bibitem{Lust:1990zi}
  D.~Lust and T.~R.~Taylor,
  Phys.\ Lett.\ B {\bf 253} (1991) 335.
  
  \bibitem{deCarlos:1991gq}
  B.~de Carlos, J.~A.~Casas and C.~Mu\~noz,
  Phys.\ Lett.\ B {\bf 263} (1991) 248.

  
  \bibitem{Vilenkin:1994pv}
  A.~Vilenkin,
  Phys.\ Rev.\ Lett.\  {\bf 72} (1994) 3137
  [arXiv:hep-th/9402085].
  \bibitem{Linde:1994hy}
  A.~D.~Linde,
  Phys.\ Lett.\ B {\bf 327} (1994) 208
  [arXiv:astro-ph/9402031].
  
  \bibitem{Linde:1994wt}
  A.~D.~Linde and D.~A.~Linde,
  Phys.\ Rev.\ D {\bf 50} (1994) 2456
  [arXiv:hep-th/9402115].
  
 
\bibitem{Sakai:1995nh}
  N.~Sakai, H.~A.~Shinkai, T.~Tachizawa and K.~i.~Maeda,
  Phys.\ Rev.\ D {\bf 53} (1996) 655
  [Erratum-ibid.\ D {\bf 54} (1996) 2981]
  [arXiv:gr-qc/9506068].

\bibitem{obserwaves}
  G.~Efstathiou and S.~Chongchitnan,
  Prog.\ Theor.\ Phys.\ Suppl.\  {\bf 163} (2006) 204
  [arXiv:astro-ph/0603118].

\bibitem{Holman:2006tm}
  R.~Holman and J.~A.~Hutasoit,
  JHEP {\bf 0608} (2006) 053
  [arXiv:hep-th/0606089].

\bibitem{Balasubramanian:2004uy}
  V.~Balasubramanian and P.~Berglund,
  JHEP {\bf 0411} (2004) 085
  [arXiv:hep-th/0408054].

\bibitem{Balasubramanian:2005zx}
  V.~Balasubramanian, P.~Berglund, J.~P.~Conlon and F.~Quevedo,
  JHEP {\bf 0503} (2005) 007
  [arXiv:hep-th/0502058].

\bibitem{Conlon:2005ki}
  J.~P.~Conlon, F.~Quevedo and K.~Suruliz,
  JHEP {\bf 0508} (2005) 007
  [arXiv:hep-th/0505076].


\bibitem{Bond:2006nc}
  J.~R.~Bond, L.~Kofman, S.~Prokushkin and P.~M.~Vaudrevange,
  arXiv:hep-th/0612197.
  
\bibitem{Ellis:2006ar}
  J.~Ellis, Z.~Lalak, S.~Pokorski and K.~Turzynski,
  JCAP {\bf 0610} (2006) 005
  [arXiv:hep-th/0606133].


  

\end{thebibliography}
\end{document}